\pdfoutput=1
\documentclass[11pt]{article}

\usepackage[final]{acl}
\usepackage{times}
\usepackage{latexsym}

\usepackage[T1]{fontenc}
\usepackage[utf8]{inputenc}

\usepackage{microtype}

\usepackage{inconsolata}

\usepackage{graphicx}

\usepackage{svg}
\usepackage{subfig}

\title{EZ-VC: Easy Zero-shot Any-to-Any Voice Conversion}

\author{Advait Joglekar, Divyanshu Singh, Rooshil Rohit Bhatia\and Srinivasan Umesh \\
        SPRING Lab, \\
        Indian Institute of Technology Madras \\ \texttt{\href{mailto:advaitjoglekar@gmail.com}{advaitjoglekar@gmail.com}}, \texttt{\href{mailto:hi.divyanshusingh@gmail.com}{hi.divyanshusingh@gmail.com}}, \\ \texttt{\href{mailto:rooshil.bhatia@gmail.com}{rooshil.bhatia@gmail.com}},  \texttt{\href{mailto:umeshs@ee.iitm.ac.in}{umeshs@ee.iitm.ac.in}}}

\begin{document}
\maketitle
\begin{abstract}
Voice Conversion research in recent times has increasingly focused on improving the zero-shot capabilities of existing methods. Despite remarkable advancements, current architectures still tend to struggle in zero-shot cross-lingual settings. They are also often unable to generalize for speakers of unseen languages and accents. In this paper, we adopt a simple yet effective approach that combines discrete speech representations from self-supervised models with a non-autoregressive Diffusion-Transformer based conditional flow matching speech decoder. We show that this architecture allows us to train a voice-conversion model in a purely textless, self-supervised fashion. Our technique works without requiring multiple encoders to disentangle speech features. Our model also manages to excel in zero-shot cross-lingual settings even for unseen languages. We provide demo samples for our model here: \href{https://ez-vc.github.io/EZ-VC-Demo/}{https://ez-vc.github.io/EZ-VC-Demo/}
\end{abstract}

\section{Introduction}

Zero-shot Voice Conversion (VC) is the task of transforming a source speaker's voice characteristics into that of a target speaker while preserving linguistic content and prosodic attributes, even for speakers unseen during training. Over the years with the advancement of modern deep learning techniques and substantial improvements in speech encoders and speech generation systems, numerous and vastly different approaches have been proposed to address this challenge. 

Textless VC architectures have become the primary area of research in this domain since cascaded ASR+TTS systems  are known to lose the non-verbal characteristics of the source speech such as laughs, whispers and other filler sounds. They also lead to cascaded errors. To overcome this, many textless VC systems these days employ either self-supervised speech encoders (SSL) or neural audio codecs (NAC) to extract speaker features or linguistic content before feeding them to a speech generation decoder. These speech representations are also often disentangled to obtain certain composite characteristics such as timbre or style. Sometimes quantized speech representations are used which form as the input for a speech generation or language model. Speech synthesis systems, which are a key component of VC architectures, have of late greatly benefited from the advancements in diffusion and continuous normalizing flow (CNF) based techniques. Voicebox\cite{le2023voiceboxtextguidedmultilingualuniversal} and its successors that use these methods are able to produce high quality audio outputs that are almost undistinguishable from real speech. These models thus show great promise for zero-shot VC tasks and yet architectures based on these methods remain under-explored. 

\noindent In this work we contribute the following,

\begin{itemize}
    \item We propose EZ-VC, a simple self-supervised any-to-any zero-shot voice conversion architecture that generalizes for unseen speakers, accents and languages while still producing highly natural and fluent speech.
    \item We demonstrate that zero-shot VC is possible without requiring multiple encoders for feature disentanglement of speaker and speech attributes.
    \item We show that combining quantized features from a self-supervised speech encoder and a flow matching speech generation decoder is sufficient to achieve state-of-the-art results.
\end{itemize}

\section{Related Work}

Early research in VC focused on disentangling speaker and content information. Works like YourTTS\cite{casanova2023yourttszeroshotmultispeakertts} focused on using speaker embeddings to extract speaker features from target speech but usually required reference text to be provided as well. Recent works like SEF-VC\cite{li2024sefvcspeakerembeddingfree} now prefer textless, speaker-embedding free VC which is also able to perform better. Since the advent of SSL speech models like Hubert\cite{hsu2021hubertselfsupervisedspeechrepresentation} and WavLM\cite{Chen_2022}, VC research has quickly learned to leverage them for their high correlation with both acoustic and linguistic content. kNN-VC\cite{baas2023voiceconversionjustnearest} works by replacing representations of source speech with the nearest neighbour from the reference speech. Vec2wav 2.0 on the other hand, uses a combination of discrete representations from vq-wav2vec for source content and WavLM features for capturing the timbre of the target speaker. At the same time, another school of approach has emerged that utilizes neural audio encoders and combines them with language models for high quality VC. Unfortunately, these systems suffer from slow inference speeds due to their auto-regressive nature. Diffusion based techniques also have been explored by DiffVC\cite{popov2022diffusionbasedvoiceconversionfast} and similar works. These models are able to demonstrate natural and robust outputs. Conditional flow-matching based speech generation methods have also begun to appear in voice conversion literature. Latest works such as AdaptVC\cite{kim2025adaptvchighqualityvoice}, StableVC\cite{yao2024stablevcstylecontrollablezeroshot}, Seed-VC\cite{liu2024zeroshotvoiceconversiondiffusion} and PFlow-VC\cite{zuo2025enhancingexpressivevoiceconversion} employ this technique for their speech decoders and generally couple them with SSL encoders. 

AdaptVC uses speaker and content encoder adapters on top of Hubert while StableVC includes three feature extractors for style, linguistic content, and mel-spectrograms. Seed-VC on the other hand requires a timber shifter module and speaker-embeddings besides a semantic feature extractor. PFlow-VC proposes a slightly different approach by using a timbre encoder for target speaker and semantic encoder for source speech. In contrast, with our architecture we wish to eliminate the need for multiple encoders or adapters for voice conversion while still being able to achieve state-of-the-art results for any-to-any VC.

% \begin{figure*}[t]
%   \centering
%   % \subfloat[Training]{\includesvg[width=0.5\textwidth]{train}}
%   \includegraphics[width=0.5\textwidth]{train.pdf}
%   % \subfloat[Inference]{\includesvg[width=0.5\textwidth]{f2}}
%   \includegraphics[width=0.5\textwidth]{f2.pdf}
%   \caption{An overview of EZ-VC}
%   \label{fig:arch}
% \end{figure*}
\begin{figure*}[t]
  \centering
  \begin{minipage}[t]{0.48\textwidth}
    \centering
    \includegraphics[width=\textwidth]{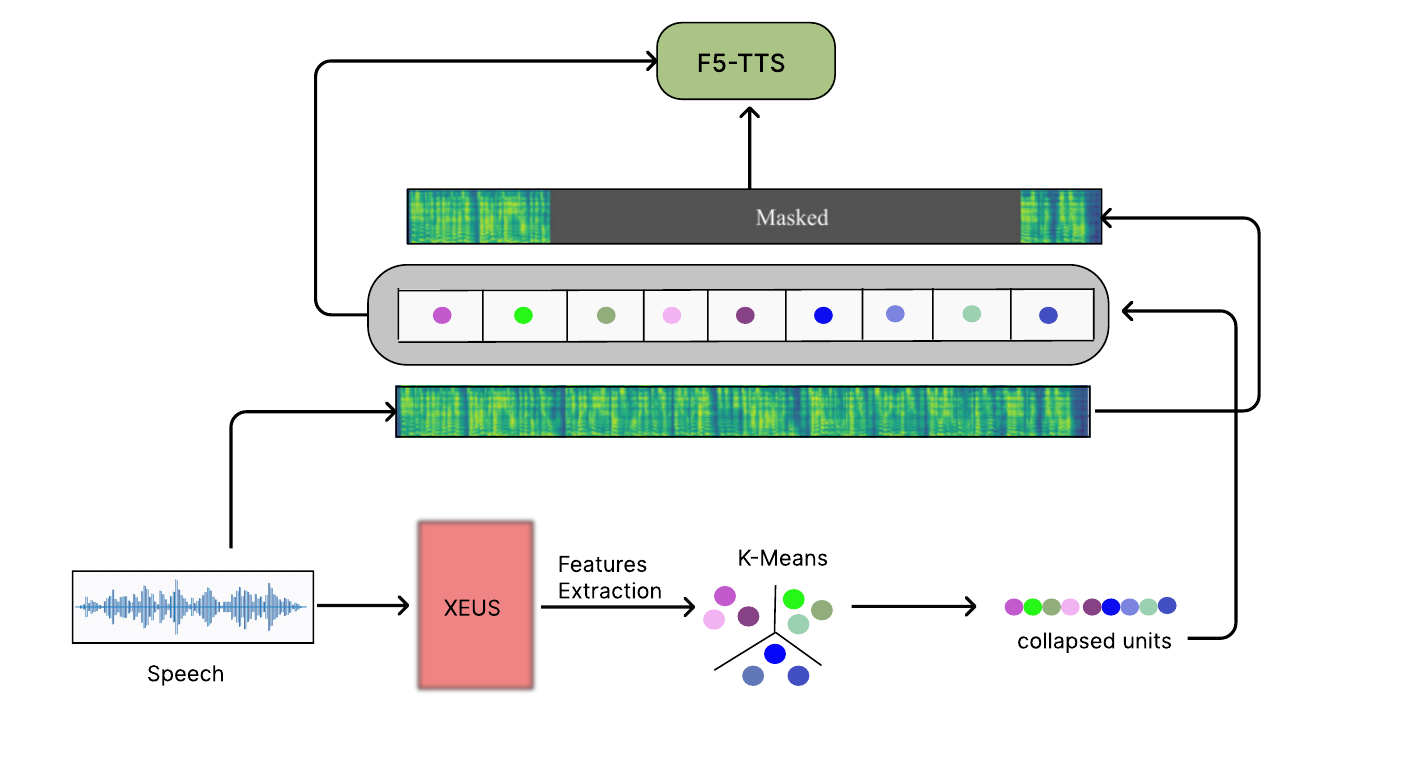}
    \textbf{(a)} Training
  \end{minipage}
  \hfill
  \begin{minipage}[t]{0.48\textwidth}
    \centering
    \includegraphics[width=\textwidth]{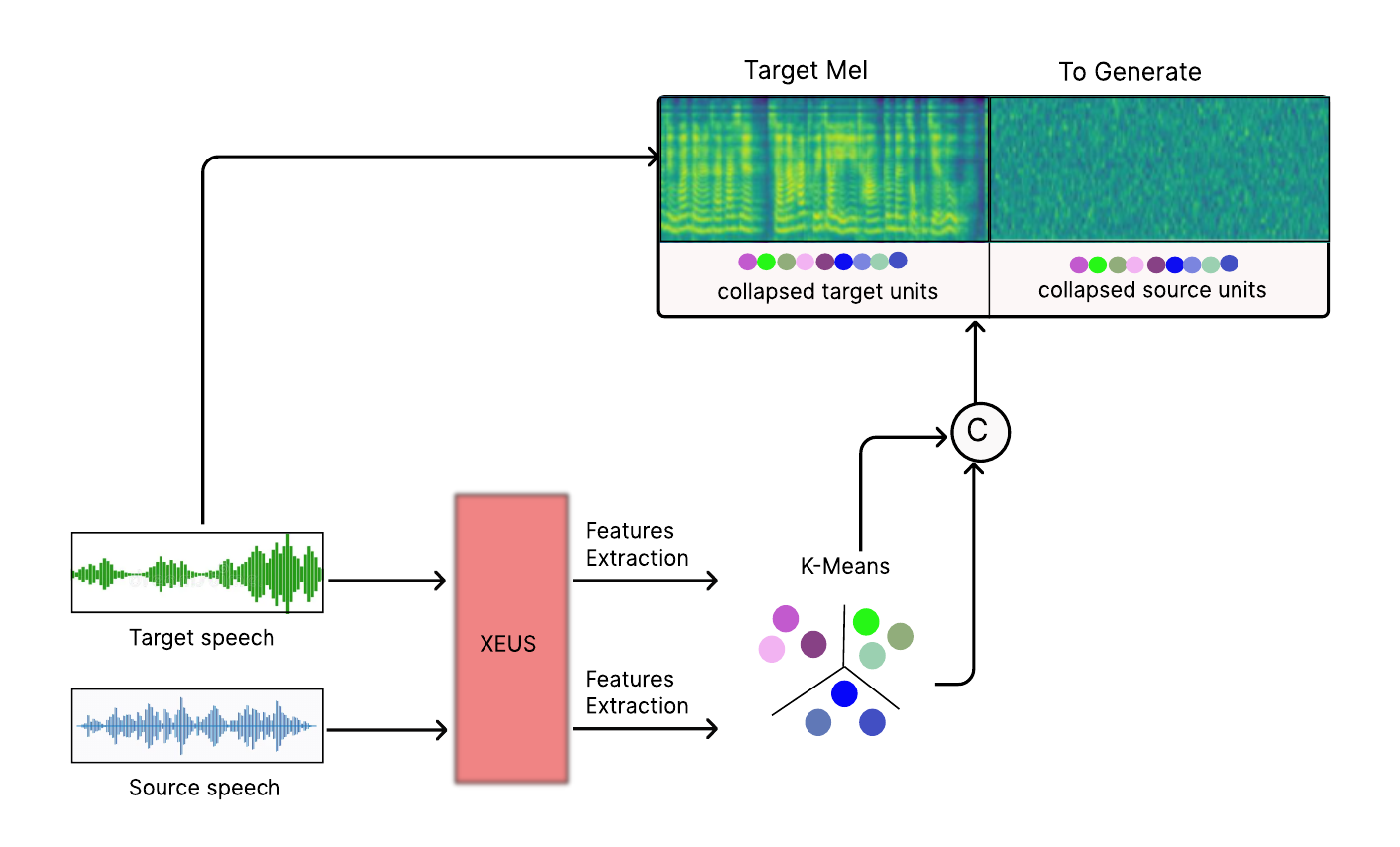}
    \textbf{(b)} Inference
  \end{minipage}
  \caption{An overview of EZ-VC}
  \label{fig:arch}
\end{figure*}

\section{EZ-VC}

EZ-VC is a simple architecture that only requires one pre-trained speech encoder and a trainable speech decoder. Unlike most other works, we do not need multiple encoders for disentanglement of speech features. Our architecture also benefits from using an off-the-shelf encoder. Other than training a simple k-means model, we do not train our speech encoding module. This helps reduce the compute and training time requirements compared to existing methods that usually ask for training both the encoder and decoder modules. 

Figure \ref{fig:arch} provides a description of our model's architecture for both training and inference. At the time of training, our model does not require any supervised or labeled data. To prepare our training set, we extract the mel-spectogram for every speech sample. These are then passed through the speech encoder first and then the resultant speech features from the 14th layer are taken and quantized using a k-means clustering model. The features are extracted at 75\% of the model depth consistent with previous works\cite{maiti2024voxtlm, communication2023seamlessm4tmassivelymultilingual}. We also de-duplicate adjacent discrete units for all samples. The mel and the corresponding discrete units become the input for our speech decoder training. With this, the model is able to learn to produce mel-spectogram from these given discrete representations and is also able to condition them based on the provided speech prompt. During inference, we pass both the source and target speech through our speech encoder system. The mel-spectogram of the target speech and its discrete units form the reference for our CFM model and the source discrete units form the prompt to generate the corresponding mel. The target and source units are concatenated and given as input to the model. The target mel is then discarded upon inference. This generated mel inherits the speaker attributes from the reference target mel while the content and style is obtained from the source units.

\subsection{Speech-to-Units}

To extract high-quality speech representations, we employ Xeus \cite{chen2024robustspeechrepresentationlearning}, a self-supervised learning (SSL) encoder trained on an extensive multilingual dataset encompassing 4,000 languages. Given its exposure to such linguistic diversity, we expect Xeus to provide robust, language-agnostic representations, enabling our model to generalize effectively to unseen languages.

Similar to WavLM, Xeus processes speech by generating frame-level embeddings. Each output embedding corresponds to a 25ms window size with a 20ms stride, effectively producing 50 embeddings per second of speech.

For the purpose of enabling speech reconstruction, we apply a quantization step using k-means clustering. Specifically, we train a 500-cluster k-means model using embeddings extracted from the 14th layer of Xeus. This clustering process provides us discrete speech units that can be used to train a units-to-speech model for resynthesis. Our k-means training dataset comprises 100 hours of English speech and 50 hours each from five Indian languages, ensuring a balanced and representative distribution of phonetic variations. This dataset is a subset of the one used for training EZ-VC.

\subsection{Units-to-Speech}

We choose the F5-TTS\cite{chen2024f5ttsfairytalerfakesfluent} architecture for our speech generation system. Building upon the work of E2-TTS\cite{eskimez2024e2ttsembarrassinglyeasy} and Voicebox, F5-TTS manages to alleviate several of their shortcomings such as duration modelling, phoneme alignment and slow convergence. We train our model for speech generation with discrete units as input. The model learns to reconstruct speech from these condensed speech representations via an infilling task. The speaker attributes are derived from the unmasked mel-spectogram and the speech content comes from the input units. This disentangles the speaker and speech, allowing us to achieve zero-shot voice conversion.

\section{Experiment}

\subsection{Datasets}

We select a wide variety of publicly available datasets for English and 5 Indian languages comprising of a total 12840 hours of speech. We hope that using a diverse set of languages and accents will help the model to generalize in unseen settings.

For English, we use 3060 hours of speech which includes a range of American, European and Indian accents. American accents come from Librispeech while European accents appear in Vox Populi\cite{wang2021voxpopulilargescalemultilingualspeech} dataset. For Indian English accent we use 1100 of speech from NPTEL\footnote{https://nptel.ac.in/} lectures.

We also select 5 Indian languages, namely Bengali, Hindi, Tamil, Telugu and Kannada to introduce diversity to our training set. We obtain in total 9780 hours of data from these languages. We procure unlabeled speech from several sources including Vaani\cite{bhogale2022effectivenessminingaudiotext}, Commonvoice\cite{ardila2020commonvoicemassivelymultilingualspeech} and datasets from IIIT-H and IIT-M. Table \ref{tab:indic} contains a full breakdown.

% Shrutilipi is a curated dataset sourced from All India Radio news bulletins and includes labeled speech data for 12 Indian languages—Bengali, Gujarati, Hindi, Kannada, Malayalam, Marathi, Odia, Punjabi, Sanskrit, Tamil, Telugu, and Urdu. The dataset comprises over 6,400 hours of speech recordings across these languages.

% Vaani is a diverse, large-scale, multimodal and multilingual dataset designed to represent India's linguistic diversity. The first phase of the dataset, covering 80 districts, consists of approximately 16,000 hours of spontaneous speech, collected from 84,600 speakers describing 130,000 images in 54 languages.

% The CommonVoice dataset is developed by Mozilla, and is an open-source, multilingual collection of voice data designed to improve speech recognition technology. It contains over 20,817 hours of validated speech data across 96 languages. We used the Tamil subset of CommonVoice 20.0 for training.

% IIIT Hyderabad, in collaboration with the Indian government's Technology Development for Indian Languages (TDIL) initiative, has undertaken a large-scale project to collect Telugu speech data. The dataset consists of 2600.8 hours of Telugu speech, collected through crowdsourcing methods.

We downsample all data, wherever neccessary to 16KHz. We further pass this data through our speech decoder combination of Xeus and k-means model to obtain discrete speech representations of each audio sample.

\begin{table*}[t]
    \centering
    \begin{tabular}{|c|c|c|c|c|}
        \hline
        & SSIM $\uparrow$ & NMOS $\uparrow$ & SMOS $\uparrow$ & UTMOS $\uparrow$ \\
        \hline
        Seed-VC & 0.69 & 3.55 & 3.78 & 3.02   \\
        kNN-VC & 0.59  & 1.94 & 2.05 & 2.42 \\
        Vec2Wav2.0 & 0.61 & 3.67 & 3.55 & 3.55 \\
        Diff-HierVC & 0.44 & 3.30 & 3.33 & 3.16 \\
        % \hline
        EZ-VC (Ours) & \textbf{0.71} & \textbf{3.91} & \textbf{3.90} & \textbf{3.56} \\
        \hline
    \end{tabular}
    \vspace{0.1cm}
    \caption{Performance metrics comparison of different VC baselines}
    \label{tab:eval}
\end{table*}

\subsection{Training setup}

We adopt the original implementation of F5-TTS for training our model. We use the base model configuration(300M params) which consists of 22 layers, 16 attention heads. For the audio samples we set sampling rate to 16KHz and use 80-dimensional log mel-filterbank features with hop length of 160. We also train a base BigVGAN\cite{lee2023bigvganuniversalneuralvocoder} model on Libri-TTS\cite{zen2019librittscorpusderivedlibrispeech} with the same configuration for a million steps. For our tokenizer, we use character level tokens with a vocabulary which includes all the 500 different discrete units.

We train this F5-TTS model from scratch with a batch size of 64 samples for 1.35 million updates on 4 NVIDIA RTX 6000 ADA GPUs. We use a peak learning rate of 5e-5 with 100k warmup steps. The rest remains the same as the original F5-TTS configuration.

\section{Evaluation}

Subjective and objective measures are equally important for evaluating voice conversion systems. In our test we use Naturalness Mean Opinion Score (NMOS) and Similarity Mean Opinion Score (SMOS) as our subjective evaluations. For objectivity, we utilize Speaker Similarity (SSim) and UTMOS\cite{saeki2022utmosutokyosarulabvoicemoschallenge} scores for comparing our models. We measure speaker similarity by using cosine similarity scores between our target speech and that of our output speech by using embeddings from a speaker verification model called ECAPA-TDNN\cite{desplanques2020ecapa}.

For our baselines, we select few of the most recent and best performing open-source voice conversion models. This makes sure that we evaluate our model against the current state-of-the-art architectures available. We select SeedVC, vec2wav 2.0, Diff-HierVC\cite{choi2023diffhiervcdiffusionbasedhierarchicalvoice} and kNN-VC as our baselines. Vec2wav and kNN-VC use primarily units-to-speech vocoders, while Diff-HierVC employs diffusion based methods. SeedVC and our work meanwhile uses CFM based speech models. 

We choose 10 samples for our evaluations. These samples are selected from various languages and accents. We prepare a variety of source and target speech combinations based on gender, inter-lingual and cross-lingual speech. We also include combinations of seen and unseen languages to test the robustness and generalization capabilities of these models. All audios are resasmpled to 16KHz to ensure fair comparison.

For our subjective evaluation, we provided these 10 samples to 20 student volunteers for comparison. Each volunteer was asked to evaluate each sample based on it's naturalness which evaluates for mainly intelligibility, style preservence, and sound quality of the output speech in comparison to the source speech. In contrast, the similarity mean opinion score judges the similarity of the speaker in the output speech to that of the target speaker. We take the average of all the samples from all the volunteers which becomes the results of our NMOS and SMOS scores.

We further objectively compare our model with Seed-VC on a seen language(English) and 2 unseen languages(German and Spanish). The results, as shown in Table \ref{tab:ssim_utmos_2}, demonstrate that EZ-VC provides better naturalness according to UTMOS, while having comparable or better speaker similarity scores. 

Analyzing the naturalness and similarity MOS scores from Table \ref{tab:eval}, we see that EZ-VC convincingly beats the latest state-of-the-art approaches for voice conversion. We find that Vec2wav 2.0, which uses discrete units coupled with a vocoder competes very well for naturalness but lags behind when it comes to imitating the target speaker. This shows that having a CFM based speech decoder is a major benefit for voice conversion systems as they are better able to capture speech styles. They also seem to generalize very well for unseen languages and accents.

\section{Conclusion}

EZ-VC hopes to make a substantial advancement in the field of zero-shot voice conversion, demonstrating that high-quality voice transformation can be achieved with a minimal architecture. By leveraging discrete speech representations from self-supervised models and a non-autoregressive speech decoder, EZ-VC balances both naturalness and speaker similarity without the need for complex feature disentanglement or multiple encoders.

 The model's ability to generalize across diverse linguistic settings highlights its robustness in cross-lingual contexts. Our findings may also suggest that discrete representations capture deeper, more universal representations of speech.

Our comprehensive evaluations show that EZ-VC achieves significantly improved capabilities for zero-shot voice conversion. We hope that our work inspires further efforts to simplify voice conversion techniques.

\section{Potential Risks}

Given the highly realistic quality of voice synthesis and the ability to achieve cross-lingual voice conversion for even unseen languages, our model carries the risk of enabling dangerous deepfakes. 

\section*{Limitations}

Despite the benifits of our approach, it has a few limitations,

\begin{itemize}
    \item The EZ-VC architecture is reliant on the quality of the pretrained speech encoder. It is likely that using an encoder trained on only one language may not achieve the level of generalization that our model does.
    \item Although our approach introduces a much simpler architecture than previous works, the computational requirements are still comparable or higher.
    
\end{itemize}

% Since December 2023, a "Limitations" section has been required for all papers submitted to ACL Rolling Review (ARR). This section should be placed at the end of the paper, before the references. The "Limitations" section (along with, optionally, a section for ethical considerations) may be up to one page and will not count toward the final page limit. Note that these files may be used by venues that do not rely on ARR so it is recommended to verify the requirement of a "Limitations" section and other criteria with the venue in question.

% \section*{Acknowledgments}

\bibliography{custom}

\begin{thebibliography}{24}
\providecommand{\natexlab}[1]{#1}

\bibitem[{Ardila et~al.(2020)Ardila, Branson, Davis, Henretty, Kohler, Meyer, Morais, Saunders, Tyers, and Weber}]{ardila2020commonvoicemassivelymultilingualspeech}
Rosana Ardila, Megan Branson, Kelly Davis, Michael Henretty, Michael Kohler, Josh Meyer, Reuben Morais, Lindsay Saunders, Francis~M. Tyers, and Gregor Weber. 2020.
\newblock \href {https://arxiv.org/abs/1912.06670} {Common voice: A massively-multilingual speech corpus}.
\newblock \emph{Preprint}, arXiv:1912.06670.

\bibitem[{Baas et~al.(2023)Baas, van Niekerk, and Kamper}]{baas2023voiceconversionjustnearest}
Matthew Baas, Benjamin van Niekerk, and Herman Kamper. 2023.
\newblock \href {https://arxiv.org/abs/2305.18975} {Voice conversion with just nearest neighbors}.
\newblock \emph{Preprint}, arXiv:2305.18975.

\bibitem[{Bhogale et~al.(2022)Bhogale, Raman, Javed, Doddapaneni, Kunchukuttan, Kumar, and Khapra}]{bhogale2022effectivenessminingaudiotext}
Kaushal~Santosh Bhogale, Abhigyan Raman, Tahir Javed, Sumanth Doddapaneni, Anoop Kunchukuttan, Pratyush Kumar, and Mitesh~M. Khapra. 2022.
\newblock \href {https://arxiv.org/abs/2208.12666} {Effectiveness of mining audio and text pairs from public data for improving asr systems for low-resource languages}.
\newblock \emph{Preprint}, arXiv:2208.12666.

\bibitem[{Casanova et~al.(2023)Casanova, Weber, Shulby, Junior, Gölge, and Ponti}]{casanova2023yourttszeroshotmultispeakertts}
Edresson Casanova, Julian Weber, Christopher Shulby, Arnaldo~Candido Junior, Eren Gölge, and Moacir~Antonelli Ponti. 2023.
\newblock \href {https://arxiv.org/abs/2112.02418} {Yourtts: Towards zero-shot multi-speaker tts and zero-shot voice conversion for everyone}.
\newblock \emph{Preprint}, arXiv:2112.02418.

\bibitem[{Chen et~al.(2022)Chen, Wang, Chen, Wu, Liu, Chen, Li, Kanda, Yoshioka, Xiao, Wu, Zhou, Ren, Qian, Qian, Wu, Zeng, Yu, and Wei}]{Chen_2022}
Sanyuan Chen, Chengyi Wang, Zhengyang Chen, Yu~Wu, Shujie Liu, Zhuo Chen, Jinyu Li, Naoyuki Kanda, Takuya Yoshioka, Xiong Xiao, Jian Wu, Long Zhou, Shuo Ren, Yanmin Qian, Yao Qian, Jian Wu, Michael Zeng, Xiangzhan Yu, and Furu Wei. 2022.
\newblock \href {https://doi.org/10.1109/jstsp.2022.3188113} {Wavlm: Large-scale self-supervised pre-training for full stack speech processing}.
\newblock \emph{IEEE Journal of Selected Topics in Signal Processing}, 16(6):1505–1518.

\bibitem[{Chen et~al.(2024{\natexlab{a}})Chen, Zhang, Peng, Li, Tian, Shi, Chang, Maiti, Livescu, and Watanabe}]{chen2024robustspeechrepresentationlearning}
William Chen, Wangyou Zhang, Yifan Peng, Xinjian Li, Jinchuan Tian, Jiatong Shi, Xuankai Chang, Soumi Maiti, Karen Livescu, and Shinji Watanabe. 2024{\natexlab{a}}.
\newblock \href {https://arxiv.org/abs/2407.00837} {Towards robust speech representation learning for thousands of languages}.
\newblock \emph{Preprint}, arXiv:2407.00837.

\bibitem[{Chen et~al.(2024{\natexlab{b}})Chen, Niu, Ma, Deng, Wang, Zhao, Yu, and Chen}]{chen2024f5ttsfairytalerfakesfluent}
Yushen Chen, Zhikang Niu, Ziyang Ma, Keqi Deng, Chunhui Wang, Jian Zhao, Kai Yu, and Xie Chen. 2024{\natexlab{b}}.
\newblock \href {https://arxiv.org/abs/2410.06885} {F5-tts: A fairytaler that fakes fluent and faithful speech with flow matching}.
\newblock \emph{Preprint}, arXiv:2410.06885.

\bibitem[{Choi et~al.(2023)Choi, Lee, and Lee}]{choi2023diffhiervcdiffusionbasedhierarchicalvoice}
Ha-Yeong Choi, Sang-Hoon Lee, and Seong-Whan Lee. 2023.
\newblock \href {https://arxiv.org/abs/2311.04693} {Diff-hiervc: Diffusion-based hierarchical voice conversion with robust pitch generation and masked prior for zero-shot speaker adaptation}.
\newblock \emph{Preprint}, arXiv:2311.04693.

\bibitem[{Communication et~al.(2023)Communication, Barrault, Chung, Meglioli, Dale, Dong, Duquenne, Elsahar, Gong, Heffernan, Hoffman, Klaiber, Li, Licht, Maillard, Rakotoarison, Sadagopan, Wenzek, Ye, Akula, Chen, Hachem, Ellis, Gonzalez, Haaheim, Hansanti, Howes, Huang, Hwang, Inaguma, Jain, Kalbassi, Kallet, Kulikov, Lam, Li, Ma, Mavlyutov, Peloquin, Ramadan, Ramakrishnan, Sun, Tran, Tran, Tufanov, Vogeti, Wood, Yang, Yu, Andrews, Balioglu, Costa-jussà, Celebi, Elbayad, Gao, Guzmán, Kao, Lee, Mourachko, Pino, Popuri, Ropers, Saleem, Schwenk, Tomasello, Wang, Wang, and Wang}]{communication2023seamlessm4tmassivelymultilingual}
Seamless Communication, Loïc Barrault, Yu-An Chung, Mariano~Cora Meglioli, David Dale, Ning Dong, Paul-Ambroise Duquenne, Hady Elsahar, Hongyu Gong, Kevin Heffernan, John Hoffman, Christopher Klaiber, Pengwei Li, Daniel Licht, Jean Maillard, Alice Rakotoarison, Kaushik~Ram Sadagopan, Guillaume Wenzek, Ethan Ye, and 49 others. 2023.
\newblock \href {https://arxiv.org/abs/2308.11596} {Seamlessm4t: Massively multilingual \& multimodal machine translation}.
\newblock \emph{Preprint}, arXiv:2308.11596.

\bibitem[{Desplanques et~al.(2020)Desplanques, Thienpondt, and Demuynck}]{desplanques2020ecapa}
Brecht Desplanques, Jenthe Thienpondt, and Kris Demuynck. 2020.
\newblock Ecapa-tdnn: Emphasized channel attention, propagation and aggregation in tdnn based speaker verification.
\newblock \emph{arXiv preprint arXiv:2005.07143}.

\bibitem[{Eskimez et~al.(2024)Eskimez, Wang, Thakker, Li, Tsai, Xiao, Yang, Zhu, Tang, Tan, Liu, Zhao, and Kanda}]{eskimez2024e2ttsembarrassinglyeasy}
Sefik~Emre Eskimez, Xiaofei Wang, Manthan Thakker, Canrun Li, Chung-Hsien Tsai, Zhen Xiao, Hemin Yang, Zirun Zhu, Min Tang, Xu~Tan, Yanqing Liu, Sheng Zhao, and Naoyuki Kanda. 2024.
\newblock \href {https://arxiv.org/abs/2406.18009} {E2 tts: Embarrassingly easy fully non-autoregressive zero-shot tts}.
\newblock \emph{Preprint}, arXiv:2406.18009.

\bibitem[{gil Lee et~al.(2023)gil Lee, Ping, Ginsburg, Catanzaro, and Yoon}]{lee2023bigvganuniversalneuralvocoder}
Sang gil Lee, Wei Ping, Boris Ginsburg, Bryan Catanzaro, and Sungroh Yoon. 2023.
\newblock \href {https://arxiv.org/abs/2206.04658} {Bigvgan: A universal neural vocoder with large-scale training}.
\newblock \emph{Preprint}, arXiv:2206.04658.

\bibitem[{Hsu et~al.(2021)Hsu, Bolte, Tsai, Lakhotia, Salakhutdinov, and Mohamed}]{hsu2021hubertselfsupervisedspeechrepresentation}
Wei-Ning Hsu, Benjamin Bolte, Yao-Hung~Hubert Tsai, Kushal Lakhotia, Ruslan Salakhutdinov, and Abdelrahman Mohamed. 2021.
\newblock \href {https://arxiv.org/abs/2106.07447} {Hubert: Self-supervised speech representation learning by masked prediction of hidden units}.
\newblock \emph{Preprint}, arXiv:2106.07447.

\bibitem[{Kim et~al.(2025)Kim, Kim, Choi, Nguyen, Mun, and Chung}]{kim2025adaptvchighqualityvoice}
Jaehun Kim, Ji-Hoon Kim, Yeunju Choi, Tan~Dat Nguyen, Seongkyu Mun, and Joon~Son Chung. 2025.
\newblock \href {https://arxiv.org/abs/2501.01347} {Adaptvc: High quality voice conversion with adaptive learning}.
\newblock \emph{Preprint}, arXiv:2501.01347.

\bibitem[{Le et~al.(2023)Le, Vyas, Shi, Karrer, Sari, Moritz, Williamson, Manohar, Adi, Mahadeokar, and Hsu}]{le2023voiceboxtextguidedmultilingualuniversal}
Matthew Le, Apoorv Vyas, Bowen Shi, Brian Karrer, Leda Sari, Rashel Moritz, Mary Williamson, Vimal Manohar, Yossi Adi, Jay Mahadeokar, and Wei-Ning Hsu. 2023.
\newblock \href {https://arxiv.org/abs/2306.15687} {Voicebox: Text-guided multilingual universal speech generation at scale}.
\newblock \emph{Preprint}, arXiv:2306.15687.

\bibitem[{Li et~al.(2024)Li, Guo, Chen, and Yu}]{li2024sefvcspeakerembeddingfree}
Junjie Li, Yiwei Guo, Xie Chen, and Kai Yu. 2024.
\newblock \href {https://arxiv.org/abs/2312.08676} {Sef-vc: Speaker embedding free zero-shot voice conversion with cross attention}.
\newblock \emph{Preprint}, arXiv:2312.08676.

\bibitem[{Liu(2024)}]{liu2024zeroshotvoiceconversiondiffusion}
Songting Liu. 2024.
\newblock \href {https://arxiv.org/abs/2411.09943} {Zero-shot voice conversion with diffusion transformers}.
\newblock \emph{Preprint}, arXiv:2411.09943.

\bibitem[{Maiti et~al.(2024)Maiti, Peng, Choi, Jung, Chang, and Watanabe}]{maiti2024voxtlm}
Soumi Maiti, Yifan Peng, Shukjae Choi, Jee-weon Jung, Xuankai Chang, and Shinji Watanabe. 2024.
\newblock Voxtlm: Unified decoder-only models for consolidating speech recognition, synthesis and speech, text continuation tasks.
\newblock In \emph{ICASSP 2024-2024 IEEE International Conference on Acoustics, Speech and Signal Processing (ICASSP)}, pages 13326--13330. IEEE.

\bibitem[{Popov et~al.(2022)Popov, Vovk, Gogoryan, Sadekova, Kudinov, and Wei}]{popov2022diffusionbasedvoiceconversionfast}
Vadim Popov, Ivan Vovk, Vladimir Gogoryan, Tasnima Sadekova, Mikhail Kudinov, and Jiansheng Wei. 2022.
\newblock \href {https://arxiv.org/abs/2109.13821} {Diffusion-based voice conversion with fast maximum likelihood sampling scheme}.
\newblock \emph{Preprint}, arXiv:2109.13821.

\bibitem[{Saeki et~al.(2022)Saeki, Xin, Nakata, Koriyama, Takamichi, and Saruwatari}]{saeki2022utmosutokyosarulabvoicemoschallenge}
Takaaki Saeki, Detai Xin, Wataru Nakata, Tomoki Koriyama, Shinnosuke Takamichi, and Hiroshi Saruwatari. 2022.
\newblock \href {https://arxiv.org/abs/2204.02152} {Utmos: Utokyo-sarulab system for voicemos challenge 2022}.
\newblock \emph{Preprint}, arXiv:2204.02152.

\bibitem[{Wang et~al.(2021)Wang, Rivière, Lee, Wu, Talnikar, Haziza, Williamson, Pino, and Dupoux}]{wang2021voxpopulilargescalemultilingualspeech}
Changhan Wang, Morgane Rivière, Ann Lee, Anne Wu, Chaitanya Talnikar, Daniel Haziza, Mary Williamson, Juan Pino, and Emmanuel Dupoux. 2021.
\newblock \href {https://arxiv.org/abs/2101.00390} {Voxpopuli: A large-scale multilingual speech corpus for representation learning, semi-supervised learning and interpretation}.
\newblock \emph{Preprint}, arXiv:2101.00390.

\bibitem[{Yao et~al.(2024)Yao, Yang, Pan, Ning, Ye, Zhou, and Xie}]{yao2024stablevcstylecontrollablezeroshot}
Jixun Yao, Yuguang Yang, Yu~Pan, Ziqian Ning, Jiaohao Ye, Hongbin Zhou, and Lei Xie. 2024.
\newblock \href {https://arxiv.org/abs/2412.04724} {Stablevc: Style controllable zero-shot voice conversion with conditional flow matching}.
\newblock \emph{Preprint}, arXiv:2412.04724.

\bibitem[{Zen et~al.(2019)Zen, Dang, Clark, Zhang, Weiss, Jia, Chen, and Wu}]{zen2019librittscorpusderivedlibrispeech}
Heiga Zen, Viet Dang, Rob Clark, Yu~Zhang, Ron~J. Weiss, Ye~Jia, Zhifeng Chen, and Yonghui Wu. 2019.
\newblock \href {https://arxiv.org/abs/1904.02882} {Libritts: A corpus derived from librispeech for text-to-speech}.
\newblock \emph{Preprint}, arXiv:1904.02882.

\bibitem[{Zuo et~al.(2025)Zuo, Ji, Fang, Jiang, Cheng, Yang, Liu, Zhang, Tu, Guo, and Zhao}]{zuo2025enhancingexpressivevoiceconversion}
Jialong Zuo, Shengpeng Ji, Minghui Fang, Ziyue Jiang, Xize Cheng, Qian Yang, Wenrui Liu, Guangyan Zhang, Zehai Tu, Yiwen Guo, and Zhou Zhao. 2025.
\newblock \href {https://arxiv.org/abs/2502.05471} {Enhancing expressive voice conversion with discrete pitch-conditioned flow matching model}.
\newblock \emph{Preprint}, arXiv:2502.05471.

\end{thebibliography}

\appendix

\section*{Appendix}
\label{sec:appendix}

\begin{table}[h]
    \centering
    \begin{tabular}{|c|c|}
        \hline
        Dataset & Hours \\
        \hline
        Librispeech & 960 \\
        Vox Populi & 1000\\
        NPTEL & 1100 \\
        \hline
        Total & 3060 \\
        \hline
    \end{tabular}
    \vspace{0.1cm}
    \caption{English Datasets}
    \label{tab:my_label}
\end{table}

\begin{table}
    \centering
    \begin{tabular}{|c|c|c|}
        \hline
        & SSIM & UTMOS \\
        \hline
      English(EZ-VC) & 87.3 & 3.76 \\
        \hline
        English(Seed-VC) & 83.9 & 3.51 \\
        \hline
    \end{tabular}
    \vspace{0.1cm}
    \caption{EZ-VC Vs Seed-VC on seen languages}
    \label{tab:ssim_utmos_1}
\end{table}

\begin{table}
    \centering
    \begin{tabular}{|c|c|c|}
        \hline
        & SSIM & UTMOS \\
        \hline
      German(EZ-VC) & 91.4 & 3.71 \\
        \hline
        German(Seed-VC) & 90.8 & 2.83 \\
        \hline
      Spanish(EZ-VC) & 84.2 & 3.49 \\
        \hline
        Spanish(Seed-VC) & 84.2 & 3.24 \\
        \hline
    \end{tabular}
    \vspace{0.1cm}
    \caption{EZ-VC Vs Seed-VC on unseen languages}
    \label{tab:ssim_utmos_2}
\end{table}

\begin{table*}
    \centering
    \begin{tabular}{|c|c|c|c|c|c|c|}
        \hline
         & Bengali & Hindi & Tamil & Telugu & Kannada & \textbf{Total} \\
         \hline
         Vaani & 1420 & - & - & 980 & 1390 & 3790 \\
         \hline
         Common Voice & - & - & 420 & - & - & 420 \\
         \hline
         Shrutilipi & 620 & - & 950 & - & - & 1570 \\
         \hline
         IIIT-H & - & - & - & 2600  & - & 2600 \\
         \hline
         IITM & -  & 1400  & - & - & - & 1400 \\
         \hline
        \textbf{Total} & \textbf{2040} & \textbf{1400} & \textbf{1370} & \textbf{3580} & \textbf{1390} & \textbf{9780} \\
         \hline
    \end{tabular}
    \caption{Indian Language Datasets}
    \label{tab:indic}
\end{table*}

\end{document}